\documentclass[useAMS,usenatbib]{mn2e}

\usepackage{graphicx}
\usepackage{amsmath}
\newcommand{\mm}{$\umu$m}
\newcommand{\ion}[2]{[#1{\sc#2}]}

\title[Cicumnuclear star formation in AGNs]{Star formation in AGNs at the
hundred parsec scale using MIR high resolution images}
\author[Daniel Ruschel-Dutra et al.]{Daniel Ruschel-Dutra$^{1,2}$\thanks{
    E-mail:daniel.ruschel@ufrgs.br (AVR)},
    Jos\'{e} Miguel Rodr\'{i}guez Espinosa$^{1}$,
\and Omaira Gonz\'{a}lez Mart\'{i}n$^{1,3,4}$, Miriani Pastoriza$^{2}$
and Rog\'{e}rio Riffel$^{2}$ \\$^1$Instituto de Astrof\'{i}sica de Canarias
\\$^{2}$Departamento de Astronomia, Instituto de F\'{i}sica da Universidade
Federal do Rio Grande do Sul\\$^3$Departamento de Astrof\'{i}sica, Universidad
de La Laguna (ULL), 38205, La Laguna, Spain\\$^4$Centro de
Radioastronom\'ia Astrof\'isica (CRyA-UNAM), 3-72 (Xangari), 8701, Morelia,
Mexico}
\begin{document}

\date{\today}

\pagerange{\pageref{firstpage}--\pageref{lastpage}} \pubyear{2016}

\maketitle

\label{firstpage}

\begin{abstract}
It has been well established in the past decades that the central black hole
masses of galaxies correlate with dynamical properties of their harbouring
bulges. This notion begs the question of whether there are causal connections
between the AGN and its immediate vicinity in the host galaxy. In this paper we
analyse the presence of circumnuclear star formation in a sample of 15 AGN
using mid-infrared observations. The data consist of a set of
11.3{\mm} PAH emission and reference continuum images, taken with ground based
telescopes, with sub-arcsecond resolution. By comparing our star formation
estimates with AGN accretion rates, derived from X-ray luminosities, we
investigate the validity of theoretical predictions for the AGN-starburst
connection. Our main results are: i) circumnuclear star formation is found, at
distances as low as tens of parsecs from the nucleus, in nearly half of our
sample (7/15); ii) star formation luminosities are correlated with the
bolometric luminosity of the AGN ($L_{AGN}$) only for objects with
$L_{AGN} \ge 10^{42}\,\,{\rm erg\,\,s^{-1}}$; iii) low
luminosity AGNs ($L_{AGN} < 10^{42}\,\,{\rm erg\,\,s^{-1}}$) 
seem to have starburst
luminosities far greater than their bolometric luminosities.
\end{abstract}

\begin{keywords}
active galactic nuclei -- interstellar medium -- star formation.
\end{keywords}

\section{Introduction}

The properties of the bulge of galaxies correlate with the properties of their
super massive black holes \citep[SMBH, ][and references therein]{Kormendy1995}.
Arguably the most widely known among these is the correspondence between the
mass ($M_\bullet$) of the SMBH and the velocity dispersion of stars in the bulge
($\sigma$), in short the $M_\bullet - \sigma$ relation
\citep{Ferrarese2000, Gebhardt2000, Gultekin2009, McConnell2013}.
Relationships such as these raise the question of whether or not there is a
causal connection between the observed properties of the active nucleus and the
host galaxy.

One of the possible physical links between the accretion-powered nuclear
activity and its immediate vicinity is the circumnuclear\footnote{Here and
throughout this paper circumnuclear refers to scales of the order of hundreds
of parsecs or less} star formation, as this implies a common mechanism for 
fuelling the growth of both the SMBH and the stellar bulge. Early works on the
subject have shown the ubiquitous presence of star formation in Seyfert
galaxies \citep{RodriguezEspinosa1987,GonzalezDelgado1993,CidFernandes2001,
Kauffmann2003,Riffel2007,Davies2007}. There has also been a number of
articles
suggesting that the feedback from accretion onto the SMBH would quench star
formation by heating the available gas, deterring the gravitational collapse of
molecular clouds \citep[and references therein]{Silk1998,Vollmer2013}. On the
other hand, the turbulence generated by supernova explosions could be
responsible for the loss of angular momentum that ultimately leads the gas to
the accretion disk \citep{Kawakatu2008,Hopkins2010,Wutschik2013}.

The MIR emission from polyciclic aromatic hydrocarbon (PAH) molecules has
frequently been used as a tracer of star formation \citep[e.g.][]{Tielens2008,
Wu2009,Gallimore2010,Diamond-Stanic2010,RuschelDutra2014}. These emission bands
are produced when UV photons from young stars heat molecules to temperatures of
the order of 1000K. The energy is subsequently radiated by fluorescence through
the many molecular modes of oscillation and vibration. Since it is unlikely
that such large molecular species would survive the intense radiation field
from the active galactic nuclei \citep[AGN, ][]{Voit1992,Siebenmorgen2004}, the
analysis of their emission is virtually free from the degeneracy considerations
which are relevant for ionic lines. However, it has been argued that the
11.3{\mm} PAH band is an indicator of recent star formation
\citep[$\sim 10^8\,\,{\rm yr}$][]{DiazSantos2010}.

This study aims at investigating the relationship between the AGN activity and
the recent star formation in its vicinity. It is only natural to search for a
causal relation between these phenomena at the smallest possible scales, e.g.
at radii of tens of parsecs from the central engine. Such regions are often
enshrouded in dusty clouds, favouring observations in wavelengths other than
the optical \citep{Nenkova2008,RamosAlmeida2011,Sales2011,GonzalezMartin2013}.
At present, ground based mid infrared (MIR) observations with 10m class
telescopes offer the possibility of obtaining high resolution images
($\sim 0.4${\arcsec}), at the same time avoiding most of the extinction from
the interstellar environment. Therefore, here we analyse high spatial
resolution PAH emission images from two similar ground-based instruments,
namely the Very Large Telescope (VLT) Imager and Spectrometer for mid Infrared
(VISIR), and the CanariCam attached to the Gran Telescopio Canarias (GTC).  

The present paper is structured as follows: in \S2 we discuss the sample
selection, observation strategy and reduction process; our estimates for star
formation rates, as well as the morphological features of the star forming
regions are examined in \S3; \S4 examines how the empirical data compare with
numerical models for the AGN-SB relation, \S5 contains a discussion of the
results and finally in \S6 we present our conclusions.

\section{The data}

\subsection{Sample selection}

The target set for this study began with the selection of 4 galaxies, three of
which were known to harbour AGNs, for observation with GTC/CanariCam, following
a criteria of spectral classification diversity. The chosen sources were the
Low Ionisation Nuclear Emission-Line Region (LINER) NGC~2146, the Seyfert 1
NGC~931 and the Seyfert 2's NGC~1194 and NGC~2273, the former being
classified as a Compton Thick source
based on X-Ray data \citep{Guainazzi2005b}.
The selected sources also followed a technical limit of detectability
for a reasonable integration time of roughly 0.2 Jy in the 12{\mm} filter of
IRAS. This is also the lower limit in the Extended 12{\mm} Galaxy Sample
\citep{Rush1993}, although NGC~2273 violates the $|b| > 25^\circ$ restriction
of this catalogue. Lastly, only galaxies with projected spatial resolutions on
the order of hundreds of parsecs were considered.

In order to add statistical weight to the present study, 12 sources observed
with VISIR at the Very Large Telescope (VLT) were added. The data for these
targets were taken from the atlas published by \citet{Asmus2014}. These
additional
galaxies consist of all the targets that are clearly detected in both the PAH2
and Si-5 filters, and respect the parameters of the last paragraph. The
similarity between the filters in CanariCam and VISIR allows for a consistent
combined analysis. The spatial scale limit applied to the proprietary data is
also respected by the VISIR observations. Basic parameters for the complete
sample of 15 AGNs are shown in table \ref{tab.obslog}.

\begin{table*}
\caption{Observation log}
\label{tab.obslog}
\begin{tabular}{l ll l c cc cc}
\hline
&&&&& \multicolumn{2}{c}{PAH 11.3} & \multicolumn{2}{c}{Si5 11.6} \\
Target & RA (J2000) & DEC (J2000) & Sp. Type & Instrument & Date Obs. & $t_{\rm exp}$ (s) &  Date Obs. & $t_{\rm exp}$ (s)\\
\hline
ESO 005-G004 & 06$^{\rm h}$ 05$^{\rm m}$ 41.7$^{\rm s}$ & -86$^{\circ}$ 37' 55.0''& Sy2      & VISIR     & 2010-11-22 & 900  & 2010-11-22 & 900  \\ 
ESO 138-G001 & 16$^{\rm h}$ 51$^{\rm m}$ 20.2$^{\rm s}$ & -59$^{\circ}$ 14' 04.2''& Sy2      & VISIR     & 2008-03-14 & 600  & 2010-07-16 & 140  \\
ESO 383-G035 & 13$^{\rm h}$ 35$^{\rm m}$ 53.8$^{\rm s}$ & -34$^{\circ}$ 17' 43.8''& Sy1.2    & VISIR     & 2004-04-14 & 180  & 2010-03-10 & 360  \\
IC 4329A     & 13$^{\rm h}$ 49$^{\rm m}$ 19.2$^{\rm s}$ & -30$^{\circ}$ 18' 33.8''& Sy1.2    & VISIR     & 2010-03-12 & 180  & 2009-05-10 & 60   \\   
IC 5063      & 20$^{\rm h}$ 52$^{\rm m}$ 02.3$^{\rm s}$ & -57$^{\circ}$ 04' 07.6''& Sy2      & VISIR     & 2006-05-05 & 180  & 2005-06-10 & 200  \\  
Mrk 1239     & 09$^{\rm h}$ 52$^{\rm m}$ 19.1$^{\rm s}$ & -01$^{\circ}$ 36' 43.5''& Sy1.5    & VISIR     & 2005-01-28 & 1000 & 2006-03-12 & 600  \\ 
NGC 253      & 00$^{\rm h}$ 47$^{\rm m}$ 33.1$^{\rm s}$ & -25$^{\circ}$ 17' 19.7''& Sy2/SB   & VISIR     & 2004-12-01 & 1500 & 2004-12-01 & 1500 \\  
NGC 931      & 02$^{\rm h}$ 28$^{\rm m}$ 14.4$^{\rm s}$ & +31$^{\circ}$ 18' 41.4''& Sy1.0    & CanariCam & 2013-09-05 & 278  & 2013-09-05 & 265  \\
NGC 1194     & 03$^{\rm h}$ 03$^{\rm m}$ 49.1$^{\rm s}$ & -01$^{\circ}$ 06' 13.0''& Sy2      & CanariCam & 2013-09-03 & 625  & 2013-09-04 & 199  \\
NGC~2146     & 06$^{\rm h}$ 18$^{\rm m}$ 37.7$^{\rm s}$ & +78$^{\circ}$ 21' 25.3''& LINER    & CanariCam & 2013-09-04 & 139  & 2013-09-03 & 662  \\
NGC 2273     & 06$^{\rm h}$ 50$^{\rm m}$ 08.6$^{\rm s}$ & +60$^{\circ}$ 50' 44.5''& Sy2ct    & CanariCam & 2013-09-08 & 625  & 2013-09-06 & 662  \\
NGC 5128     & 13$^{\rm h}$ 25$^{\rm m}$ 27.6$^{\rm s}$ & -43$^{\circ}$ 01' 08.8''& Sy2      & VISIR     & 2006-04-09 & 180  & 2006-03-15 & 600  \\
NGC 5506     & 14$^{\rm h}$ 13$^{\rm m}$ 14.9$^{\rm s}$ & -03$^{\circ}$ 12' 27.2''& Sy1.9    & VISIR     & 2010-02-23 & 180  & 2006-06-06 & 600  \\
NGC 5995     & 15$^{\rm h}$ 48$^{\rm m}$ 24.9$^{\rm s}$ & -13$^{\circ}$ 45' 28.0''& Sy2/SB   & VISIR     & 2010-07-26 & 1000 & 2010-07-26 & 1000 \\
NGC 6240     & 16$^{\rm h}$ 52$^{\rm m}$ 58.8$^{\rm s}$ & +02$^{\circ}$ 24' 03.6''& Sy2/LINER    & VISIR     & 2005-04-19 & 1800 & 2005-04-19 & 1800 \\
NGC 7469     & 23$^{\rm h}$ 03$^{\rm m}$ 15.6$^{\rm s}$ & +08$^{\circ}$ 52' 25.3''& Sy1.2    & VISIR     & 2006-07-12 & 180  & 2006-06-15 & 600  \\
\hline
\end{tabular}
\end{table*}

Our sample of nearby AGN is far from being complete both in volume and
luminosity. A cross match between the AGN catalogue of \citet{VeronCetty2010}
and the Extended 12~{\mm} Galaxy Sample \citep{Rush1993} returns 44 sources
with $z < 0.03$ and spectral classification of either Seyfert or LINER. If we
extrapolate on the region close to the galactic equator left out from
\citet{Rush1993} there should be $\sim76$ galaxies that fit the description.
Consequently, our sample represents close to 20\% of the AGNs with $z <0.03$
and $F_{12} > 0.2$~Jy.

\subsection{Observations and data reduction}
\label{sec.redux}

\begin{figure}
\centering
\includegraphics[width=\columnwidth]{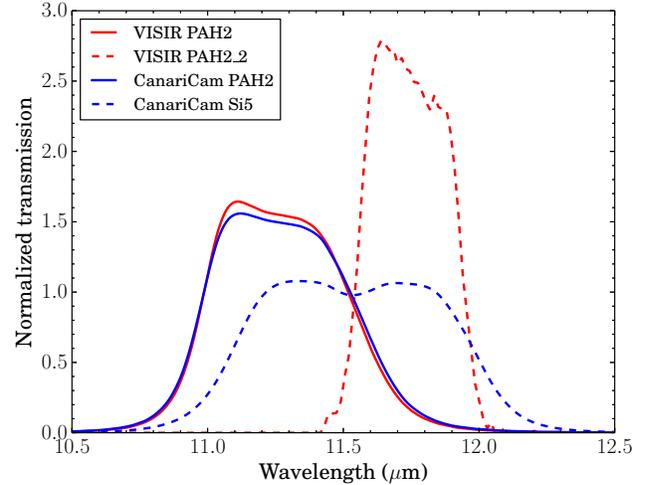}
\caption{
    Transmission profiles of the VISIR (\emph{red}) and CanariCam
    (\emph{blue}) filters. Solid and dashed lines represent the emission and
    continuum filters respectively, with respect to the rest frame wavelength.
    The transmission integral of each filter has been normalised.}
\label{fig.filterprofiles}
\end{figure}

New proprietary data was acquired with CanariCam in the filters PAH2
($\lambda_{\rm c} = 11.26$~{\mm}) and Si5 ($\lambda_{\rm c} = 11.53$~{\mm}).
Additional archival data consists of VISIR observations with filters PAH2
($\lambda_{\rm c} = 11.26$~{\mm}) and PAH2\_2 ($\lambda_{\rm c} = 11
.73$~{\mm}). This set of four filters, a pair for each instrument, was chosen
to yield the closest possible to continuum free PAH emission images. The
transmission profiles of all the filters employed in this work are shown in
figure \ref{fig.filterprofiles}, where the transmission coefficient was
normalised so that the total area under each curve equals unity.

The field of view for CanariCam is a rectangle measuring 
26{\arcsec} x 19{\arcsec},
with a spatial sampling of 0.08\arcsec/pixel. For VISIR, however, we used data
from two distinct observation modes, with square fields of view of sides
32.5{\arcsec} and 19.2{\arcsec}.
The corresponding pixel scales are 0.127\arcsec/pixel
and 0.075\arcsec/pixel.

The CanariCam data acquisition followed the standard recipe for MIR ground
based observations, with thermal emission from the telescope and the atmosphere
being removed by the chopping/nodding technique. The same applies to the
archival VISIR data. In the latter case, some of the nod frames which had
intense detector artefacts were classified by visual inspection and discarded,
resulting in a total exposure time slightly lower than the one originally
intended.


The reduction process for the CanariCam data employed the RedCan pipeline
\citep{GonzalezMartin2013}, with the addition of a recently developed algorithm
to register the centroid in a sequence of nod images and realign the exposures.
This allowed us to obtain nearly diffraction limited images from what was
otherwise a seeing limited stack. Standard stars, observed not more than an
hour apart from the science exposures, were used to flux calibrate the images.
The VISIR images were used as published in \citet{Asmus2014}, thus we refer the
reader to that paper for details about the reduction process.


\section{Imaging Analysis}

In this section we discuss the methodology and results from the analysis of the
MIR images. At first the problem of isolating the location of PAH emission is
investigated, followed by a description of the new observations with CanariCam.
Finally photometric measurements are discussed.

\subsection{PAH emission maps}
\label{sec.pahmaps}

The simplest approach to the problem of isolating the 11.3~{\mm} PAH emission
would be to produce maps, employing a method to compensate for the different
continuum levels in both filters. The objective here is not yet to reach an
accurate measurement of PAH emission, but rather to locate excesses in the PAH
sampling filter with respect to the expected continuum level.

The width of MIR filters tends to be considerably larger than their
counterparts in the optical range, close to half a micrometer, as can be seen
in figure \ref{fig.filterprofiles}. With central wavelengths separated by
distances that may be smaller than the filter's full width at half maximum
(FWHM), it is clear that there is considerable overlap between the emission and
reference filters in both instruments. This is particularly true in the case of
CanariCam, where the continuum reference filter has nearly 60\% of its
transmission curve in common with the PAH2 filter. Furthermore, the distance
between the filters, although small in comparison to their FWHM, are large
enough for differences in continuum levels to become noticeable. Therefore, a
simple subtraction of the continuum level, as probed by the reference filter,
cannot lead to an accurate estimate of the PAH emission.

Analysing tens of AGNs with already published high resolution MIR spectra
\citep[e.g.][]{RuschelDutra2014,Sales2014,GonzalezMartin2013,Esquej2014}, one
finds that low resolution spectra, taken with the Infrared Spectrometer (IRS)
aboard the Spitzer space telescope, show an underlying continuum that is very
similar to that of the high resolution spectra. The reason for this agreement
lies in the dominant role of power law and warm dust emission associated with
the AGN, plus the silicate absorption bands at 9.7~{\mm} and 16.8~{\mm}. The
later, even if not directly linked to the active nucleus, imprints its effect
on the line of sight that leads to the AGN.

In order to produce a model of the continuum we examined the \emph{Spitzer}/IRS
spectra of our targets, available at the \emph{Spitzer} archive
\footnote{http://sha.ipac.caltech.edu/applications/Spitzer/SHA/}, with the
spectral analysis tool {\sc pahfit} \citep{Smith2007a}. By fitting the spectrum
as a combination of continuum emission, silicate absorption, and emission from
molecules and ions, the code is able to return a sophisticated estimate for the
continuum shape. Since we are studying galaxies, the redshift of the spectrum
had to be taken into consideration when producing the model. The reader should
keep in mind that \emph{Spitzer}/IRS has a slit width of 3.6\arcsec, and
therefore is sampling a much larger region of the galaxy than ground based
observations. Nevertheless the dominance of the nucleus over the host galaxy
makes it possible to obtain a fair approximation of the continuum using low
resolution spectroscopy.


We use this continuum emission as an estimate of the slope of the continuum
under the PAH feature. Assuming that the same continuum shape holds throughout
the image's field of view, we can then produce a qualitative map of the
deviation from the null hypothesis, which is having no PAH emission at all. In
other words, we are building a qualitative map of PAH ``excess''. The operation
can be described by the equation

\begin{equation}
    E = \left| I_1 - I_2 \frac{\int F_1 C\,{\rm d}\lambda}
        {\int F_2 C\,{\rm d}\lambda} \right|
\label{pahexcess}
\end{equation}

\noindent where $E$ is the difference from the null hypothesis, $I_n$ is the
image in the $n$ filter, $F_n$ is the normalised transmission function of the
$n$ filter and $C$ is the function describing the semi-empirical continuum.
This model is limited by the possibility of spatial variations of the continuum
function, which could also produce an ``excess'' as the one described above.
Nevertheless, there is currently no better alternative to infer the continuum
slope. We would also like to emphasise that the continuum function is
independently modelled for each target according to its \emph{Spitzer}/IRS
spectrum and redshift.

\begin{figure}
\includegraphics[width=\columnwidth]{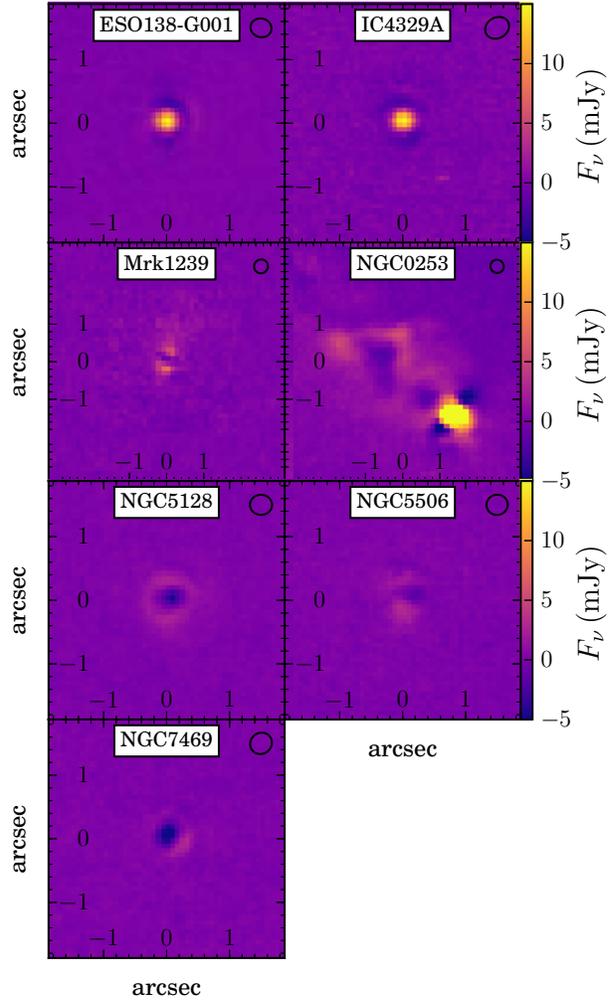}
\caption{
    Continuum subtracted 11.3{\mm} PAH images for the galaxies classified as
    displaying resolved or unresolved emission. The ellipses in the upper right
    corner of each image represent the FWHM of the standard star.}
\label{fig.subpontual}
\end{figure}

Employing this method we conclude that from the 15 galaxies in our sample only
four show PAH emission in the image subtraction.
Of these, only NGC~253
displays an extended structure, with all the other galaxies appearing as
unresolved sources. The PAH emission maps that resulted from the continuum
subtraction are displayed in figure \ref{fig.subpontual}, for the spatially
unresolved and resolved sources. Notably, as can be seen in table
\ref{tab.obslog}, the images on each filter have not been taken in the same
observing night for all the targets. Therefore, the point spread function (PSF)
of the images are not naturally matched. Perhaps the most notable case of PSF
mismatch are the observations of Mrk~1239, which show lobes extending beyond
the FWHM of the central source. We thus refrain from assertions on morphology
features that are below the wider of the two PSFs. Considering this caveat,
only NGC~253 can be safely classified as having extended PAH emission.

Apart from the four galaxies with larger fluxes on the PAH images, we find two
sources, namely NGC~5128 and NGC~7469, which have fluxes in the PAH images
lower than the expected from the inferred continuum slope. This effect can be
due to an overestimate of this slope, or to some of the emission from the
molecular band ``leaking'' into the reference filter. This last issue is
further discussed in section \ref{sec.sblum} and in the appendix,
where we present our method for dealing with the effects of redshift in the
relative fluxes between the filters. Additionaly, NGC~5506 shows
a small, almost unresolved, PAH ``excess'' and a region where the continuum
image is more intense than expected, leading to slightly negative values in
figure \ref{fig.subpontual}. Since the net result is close to zero, we
conservatively exclude this galaxy from the list of positive detections.

It is important to keep in mind the projected scale of these images, and
consequently the maximum radius of an unresolved source. For Mrk~1239, the
farthest galaxy in figure \ref{fig.subpontual}, the PAH emitting region is at
most 400~pc away from the central engine, and for the nearest target, namely
NGC~253, the same region is no further than 19~pc. In the case of NGC~253,
structures seen in emission are coincidental with H II regions already
identified in the literature \citep{Forbes2000,Lira2007}.


\subsection{Proprietary CanariCam images}
\label{sec.proprietary}

\begin{figure}
\includegraphics[width=\columnwidth]{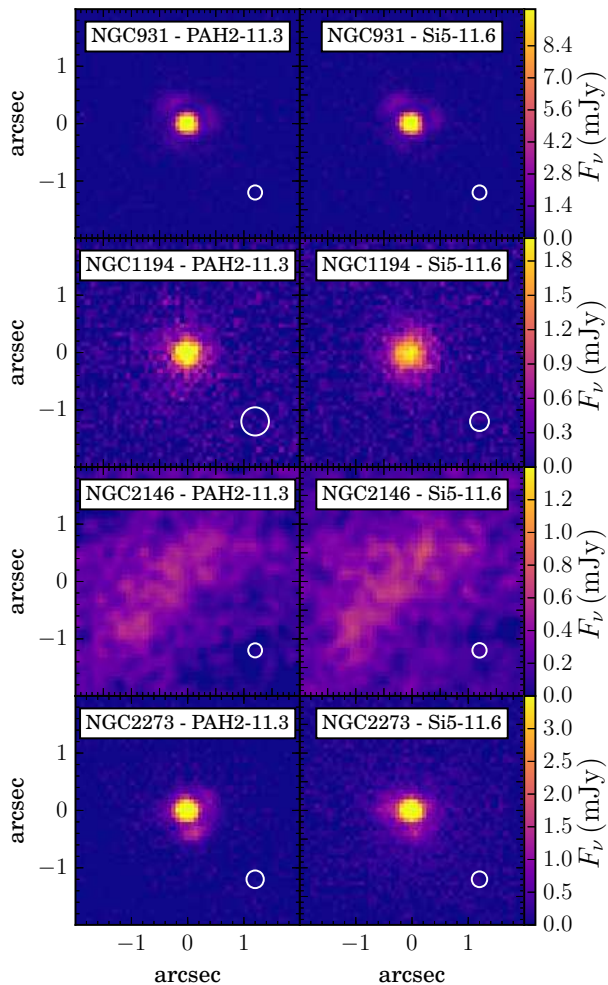}
\caption{
    Images acquired with CanariCam. With the exception of NGC~2146, for which
    no clear nucleus can be identified, all the other galaxies in this set
    appear as point like sources. In order to highlight the extended structure
    of NGC~2146 its image has been convoluted by a Gaussian. The faint ring
    like structures seen in NGC~931 and NGC~2273 are very likely instrumental
    artefacts (see text). In all figures North is up, East is left, and the
    ellipse in the lower right represents the FWHM of the standard star.}
\label{fig.canaricamimages}
\end{figure}

Since some of our targets are having their MIR images published for the first
time in this paper, we take this opportunity to examine their images in more
depth. Four galaxies were observed with CanariCam: NGC~931, NGC~1194, NGC~2146
and NGC~2273. The images in the filters PAH2 - 11.3~{\mm} and Si5 - 11.6~{\mm}
for the four galaxies are shown in figure \ref{fig.canaricamimages}. For
display purposes, the images of NGC~2146 have been convolved with a Gaussian
with $\sigma$ of one pixel, in order to emphasise the large scale structure.

At this depth and spatial resolution, the galaxies which harbour an AGN appear
as point-like sources in both filters, including the relatively close NGC~2273
($z = 0.006$). In contrast, NGC~2146 shows diffuse emission in the form of a
band extending from southeast to northwest. The direction of this structure
coincides with the dense dust lane identifiable in optical and near infrared
images of this galaxy \citep[e.g., ][]{Martini2003}. Although this galaxy is
classified as a LINER, no clear nucleus could be identified, and therefore it
was left out of the AGN sample. We conclude that this galaxy probably has a
LINER-like emission attributable to sources other than the SMBH accretion disk,
such as shocks or evolved stars \citep{Filippenko1984,Stasinska2008}.

Only one of these four sources shows an appreciable difference in flux between
filters, namely the Seyfert~2 (Sy2) NGC~1194. The similarity in filter fluxes
is an expected result, since the transmission curves overlap and the distance
between central wavelengths is comparable to the FWHM of the emission band we
are trying to probe.

\subsection{Artefacts in CanariCam imaging}

In some of the images, particularly in the case of NGC~931 and NGC~2273, one
can clearly see a pattern of three bright spots circling the central object.
Since the same pattern is also visible in the standard star images they are
almost certainly not real. This pattern is a combination of several effects:
the hexagonal shape of GTC's mirror and its segments; a small difference in
phase between segments; and small guiding errors. The effect is not apparent in
the images of NGC~1194 due to its lower signal to noise ratio.

The last of the above mentioned issues is related to the telescope guiding and
the way CanariCam records the data. Each saveset is the result of nearly six
seconds of chop cycles, which are stored in a buffer before being written as a
file. The source position between frames varies as much as $\sim 0.8${\arcsec}
peak-to-peak, and in fact it is apparent by the structures seen in each of the
savesets that such variations also occur between chop cycles. Registered
stacking of the accumulated chop frames can easily solve the problem of image
movement between savesets, as long as the targets are bright enough to be
detected in each frame. Nevertheless, movement within each saveset requires a
guiding correction frequency at least equal to the chopping frequency.

Unlike the diffraction features, which are directly related to the relative
position between the detector and the primary mirror, the guiding artefacts are
stochastic in nature. Thus it is impossible to eliminate them by simple
comparison with the standard star. These effects, while potentially harmful to
the morphological analysis, are in now way detrimental to the photometry,
provided that the apertures include the stray light. 

\subsection{Photometry}
\label{sec.photo}

Fluxes were obtained from the images through aperture photometry, considering
virtual pupils with radii equal to the FWHM of the standard star employed in
the flux calibration. The background levels were evaluated from an annulus with
width equal to the aperture's radius, separated from the latter by half its
radius. All the sums were performed by our own routines which include treatment
of partial pixels. Photometric data are displayed in table \ref{tab.phot}.

Signal to noise ratio estimates for MIR images have a few differences from
their optical counterparts. The consideration that atmospheric emission follows
a Poisson distribution does not hold for a chop frequency of a few tens of
milliseconds. The reason is that at such short intervals there is a significant
correlation between background levels in subsequent frames. Moreover, the
background level is also a function of the emission from the telescope itself,
making it even more time dependent. We chose to estimate the noise levels from
the standard deviation of the background in the annulus.

\begin{table*}
\renewcommand{\tabcolsep}{1.3mm}
\caption{Photometric fluxes}
\label{tab.phot}
\begin{tabular}{c l cccccccccc}
\hline
ID & Target & z & $L_{\rm 2-10~keV}$ & Aperture & $f_{\rm PAH}^1$ &
    $f_{\rm REF}^1$ & $L_{\rm AGN}$ & $L_{\rm PAH}$ & $L_{\rm SB}$ & SFR\\
\\
&& & $(\log {\rm erg\,\,s}^{-1})$ & (parsec) & (mJy) & (mJy) &
    $(\log {\rm erg\,\,s}^{-1})$ &  $(10^{40}\,\,{\rm erg\,\,s^{-1}})$ &
    $(\log {\rm erg\,\,s}^{-1})$ & (${\rm M}_\odot\,\,{\rm yr}^{-1}$) \\
\\
\hline
1 & ESO005-G004   & 0.006 & 41.92$^{a}$   & 128.7 &$   19 \pm   4$ & $   24 \pm   5$ & 42.94 & $    13^{+  28}_{   -9}$ & $  43.3^{+ 0.5}_{ -0.5}$ & $   0.3^{+ 0.7}_{ -0.2}$\\
2 & ESO138-G001   & 0.009 & 42.52$^{b}$   & 191.0 &$  665 \pm  11$ & $  640 \pm  14$ & 43.65 & $    64^{+ 137}_{  -43}$ & $  43.9^{+ 0.5}_{ -0.5}$ & $   1.6^{+ 3.4}_{ -1.1}$\\
3 & ESO383-G035   & 0.008 & 42.41$^{c}$   & 164.0 &$  305 \pm   8$ & $  301 \pm  19$ & 43.52 & $<\,    4$ & $<\, 42.8$ & $<\,$   0.1\\
4 & IC4329A       & 0.016 & 43.70$^{d,e}$ & 332.2 &$  973 \pm  12$ & $  905 \pm  25$ & 45.13 & $   855^{+1849}_{ -585}$ & $  45.1^{+ 0.5}_{ -0.5}$ & $    21^{+  46}_{  -15}$\\
5 & IC5063        & 0.011 & 42.94$^{f}$   & 234.6 &$  616 \pm  14$ & $  809 \pm  17$ & 44.17 & $<\,   19$ & $<\, 43.4$ & $<\,$   0.5\\
6 & Mrk1239       & 0.020 & 40.82$^{g}$   & 413.2 &$  602 \pm   9$ & $  522 \pm   8$ & 41.70 & $<\,  302$ & $<\, 44.6$ & $<\,$   7.5\\
7 & NGC0253       & 0.001 & 40.00$^{h}$   &  18.7 &$ 1768 \pm  97$ & $ 1561 \pm 114$ & 40.83 & $   7.8^{+16.8}_{ -5.3}$ & $  43.0^{+ 0.5}_{ -0.5}$ & $   0.2^{+ 0.4}_{ -0.1}$\\
8 & NGC0931       & 0.016 & 43.28$^{i}$   & 338.5 &$  210 \pm  10$ & $  217 \pm  12$ & 44.60 & $<\,  540$ & $<\, 44.9$ & $<\,$  13.5\\
9 & NGC1194       & 0.014 & 42.64$^{j}$   & 282.4 &$  176 \pm  17$ & $  167 \pm   7$ & 43.80 & $<\,  211$ & $<\, 44.5$ & $<\,$   5.3\\
10 & NGC2273      & 0.006 & 42.23$^{k}$   & 128.7 &$  111 \pm   5$ & $  106 \pm   6$ & 43.30 & $    25^{+  55}_{  -17}$ & $  43.5^{+ 0.5}_{ -0.5}$ & $   0.6^{+ 1.4}_{ -0.4}$\\
11 & NGC5128      & 0.002 & 42.31$^{l}$   &  37.4 &$  823 \pm  17$ & $  991 \pm  13$ & 43.40 & $   1.3^{+ 2.8}_{ -0.9}$ & $  42.3^{+ 0.5}_{ -0.5}$ & $  0.03^{+0.07}_{-0.02}$\\
12 & NGC5506      & 0.006 & 43.01$^{c}$   & 122.5 &$  704 \pm  15$ & $  839 \pm  13$ & 44.26 & $    13^{+  28}_{   -9}$ & $  43.2^{+ 0.5}_{ -0.5}$ & $   0.3^{+ 0.7}_{ -0.2}$\\
13 & NGC5995      & 0.025 & 43.53$^{m}$   & 521.2 &$  290 \pm   3$ & $  291 \pm   5$ & 44.92 & $<\,   23$ & $<\, 43.5$ & $<\,$   0.6\\
14 & NGC6240      & 0.024 & 44.26$^{n}$   & 504.6 &$  152 \pm  15$ & $  185 \pm  12$ & 45.87 & $<\,  116$ & $<\, 44.2$ & $<\,$   2.9\\
15 & NGC7469      & 0.016 & 43.02$^{o}$   & 330.1 &$  393 \pm  11$ & $  479 \pm   9$ & 44.27 & $<\,   68$ & $<\, 44.0$ & $<\,$   1.7\\
\hline
\end{tabular}

{\small $^1$ These flux densities are the direct result of aperture photometry
    over the monochromatic images, without any of the procedures for continuum
    subtraction or redshift compensation discussed in the text. References for
    the X-ray luminosities: $^a$ \citet{Winter2009};
$^b$~\citet{Piconcelli2011};
$^c$~\citet{Nandra2007};
$^d$~\citet{Shinozaki2006};
$^e$~\citet{Bianchi2009};
$^f$~\citet{Marinucci2012};
$^g$~\citet{Corral2011};
$^h$~\citet{MullerSanchez2010};
$^i$~\citet{Ueda2011};
$^j$~\citet{Greenhill2008};
$^k$~\citet{Awaki2009};
$^l$~\citet{Shu2011};
$^m$~\citet{Ueda2005};
$^n$~\citet{GonzalezMartin2006};
$^o$~\citet{GonzalezMartin2012};
$^p$~\citet{Brightman2011};  }
\end{table*}

\begin{figure}
\includegraphics[width=\columnwidth]{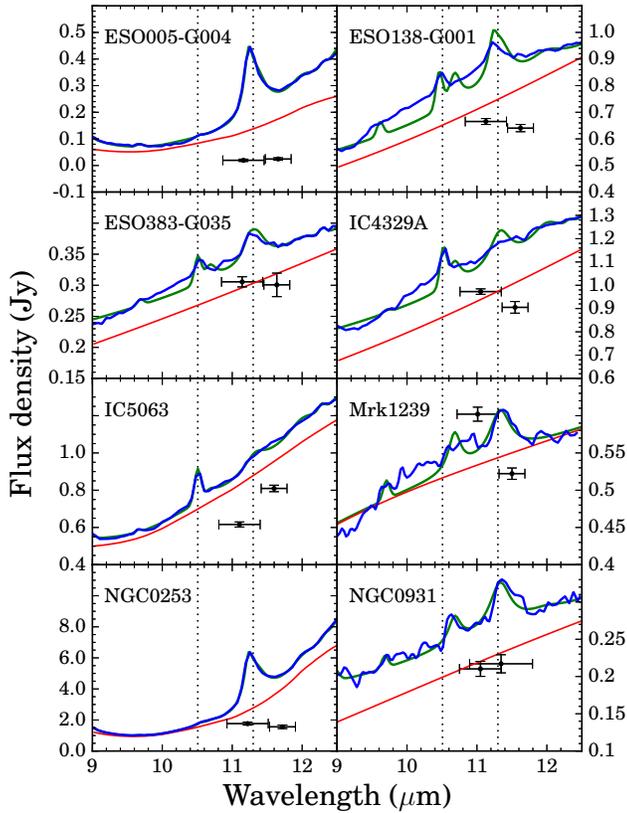}
\caption{
    Spectra from \textit{Spitzer}/IRS for the first 8 galaxies in the
    sample. Blue lines are the observed spectra shifted to the rest
    frame; red lines represent the continuum emission with the silicate
    absorption as considered by {\sc pahfit}; vertical dotted lines
    mark the wavelength of the 10.5 {\mm} [S {\sc iv}] line and 11.3 {\mm} PAH
    band; green lines represent the best theoretical match to the
    spectrum. The dots are the photometric flux points shifted to the central
    wavelength they are sampling at the given redshift, with error bars
    representing the filter's FWHM and uncertainties in the flux measurements,
    in the horizontal and vertical directions respectively.}
\label{fig.spectra}
\end{figure}

\begin{figure}
\includegraphics[width=\columnwidth]{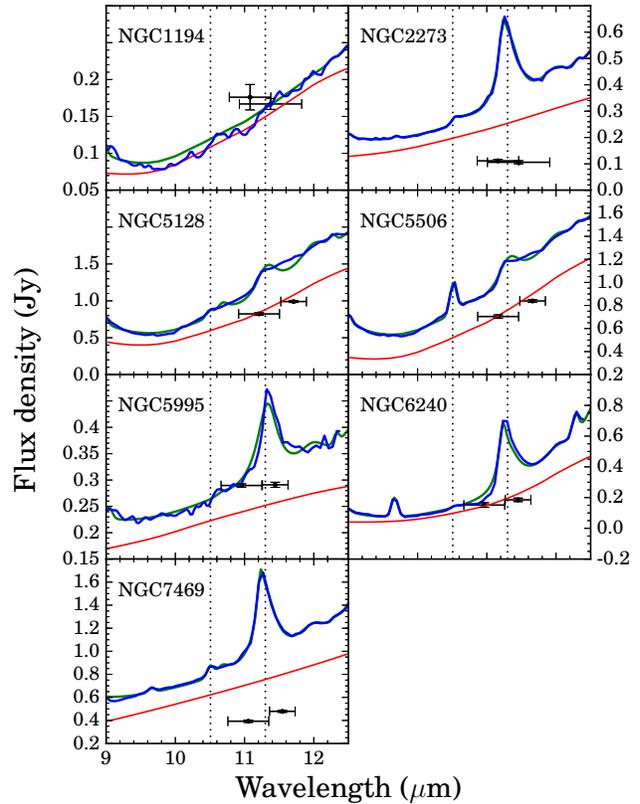}
\caption{Same as figure \ref{fig.spectra} for galaxies 9 through 15.}
\label{fig.spectra2}
\end{figure}

In figures \ref{fig.spectra} and \ref{fig.spectra2} we present the photometric
points for the 15 galaxies in the sample, plotted along archival spectra from
\emph{Spitzer/IRS}. Since the spatial resolution of \emph{Spitzer} is much
lower than that of the data of CanariCam and VISIR, the flux in the nuclear
extractions of the later is naturally smaller. We also show in these figures
the continuum inferred from the spectral fitting with {\sc pahfit}. 

The photometry of all the other galaxies seems to be in good agreement with the
\emph{Spitzer/IRS} spectra, in the sense that the photometric points are at
most equal to the spectroscopic flux. Interestingly, galaxies showing
prominent 10.5{\mm} \ion{S}{iv} emission are the ones that show better
agreement between nuclear photometry and host galaxy spectrum, probably due to
the prevalence of the central source. Whereas in galaxies with weak \ion{S}{iv}
emission the AGN is correspondingly less dominant in the low spatial resolution
spectrum. Targets showing very strong PAH emission in the \emph{Spitzer/IRS}
spectra tend to have nuclear fluxes well below those of the host galaxy, and
even below {\sc pahfit}'s estimate for the continuum.

\section{Star formation rates}
\label{sec.sblum}

In order to estimate the circumnuclear star formation rates based on the MIR
photometry, we developed a method based on the difference between the
photometric fluxes already discussed and careful measurements of the
continuum slope. We
stress that given the different redshifts of the galaxies and the large
variations in silicate profile around 10{\mm}, it is impossible to determine
the flux in PAH emission features based solely on the photometric fluxes.
We refer the reader to the appendix \ref{sec.diff2flux} for the technical
details of this analysis.


Our results, presented in table \ref{tab.phot}, show that 7 out
of the 15 galaxies in the sample have
PAH emission. This number is different from the one presented in section
\ref{sec.pahmaps} due to the different approach. Most importantly, simulating
the PAH emission band provides information on the effects of having some
of the emission sampled by the reference filter.
Also, while in section \ref{sec.pahmaps} the images were compared on a pixel
by pixel basis, in the present analysis only the integrated photometry is
considered, thus rendering PSF mismatch problems irrelevant.
Therefore, not all galaxies identified in the image subtraction resulted in
positive detections of PAH and \emph{vice-versa}.

Once known, the total flux emitted in the 11.3~{\mm} PAH flux can be used in
onjunction with empirical relations for the SFR. Although this molecular
feature has a less stringent correlation with SFR than the emission band at
8.6~{\mm} \citep{Diamond-Stanic2010}, the features at 8.6~{\mm} appear to be
observationally suppressed in the vicinity of the AGN, while the 11.3~{\mm}
seems to be less affected \citep{Alonso-Herrero2011}. That apparent suppression
should not be confused with the physical destruction of PAH molecules, but
rather the dilution of PAH emission in the intense continuum emission from the
AGN \citep{Sales2010,Alonso-Herrero2014}.

As previously discussed in the introduction, the 11.3~{\mm} PAH band is a
reliable MIR proxy for star formation rates. It is particularly well suited for
the study of AGN due to its relative insensitivity to the radiation from the
accretion disk, which is thought to destroy the molecules \citep{Voit1992}.
This is not the case of atomic fine structure lines such as the 12.8~{\mm}
\ion{Ne}{ii}, which tend to overestimate SFR in luminous AGNs
\citep{Diamond-Stanic2010}. Other PAH features on shorter wavelengths, such
as the 6.2, 7.7 and 8.6~{\mm} features, have been shown to be suppressed in
Seyferts (\emph{ibid.}).

Based on templates of MIR spectra of starburst galaxies from \citet{Rieke2009},
\citet{DiamondStanic2012} derived the relation

\begin{equation}
    \dot{M}_*\,\,(M_\odot\,\,{\rm yr^{-1}}) = 9.6 \times 10^{-9} L_{\rm PAH}
    (L_\odot)
\label{sfrpah}
\end{equation}

\noindent for starbursts characterised by
$\dot{M}_* < 10\,\,M_\odot\,\,{\rm yr^{-1}}$, where $L_{\rm PAH}$ is the
luminosity from the PAH 11.3~{\mm} band. This equation has been evaluated for
galaxies with $10^{9.75} < L_{\rm IR} < 10^{10.75}$, with a dispersion of
0.28~dex. Our sample has a median $L_{\rm IR} = 10^{10.65}$ with three galaxies
above $L_{\rm IR} = 10^{11}$, therefore we conservatively regard uncertainties
in $\dot{M}_*$ as 1 dex.

Proceeding in this line of thought, the energy-mass equivalence allows the
luminosity of the circumnuclear starburst ($L_{\rm SB}$) to be expressed as
function of the mass conversion rate from molecular gas to stars ($\dot{M}_*$).
Adding a scaling factor of 0.14 \citet{Kawakatu2008} write $L_{\rm SB}$, at a
particular instant, as

\begin{equation}
    L_{\rm SB} = 0.14\varepsilon \dot{M}_* c^2
\end{equation}

\noindent where $c$ is the speed of light in the vacuum and
$\varepsilon$ is the mass conversion efficiency set to 0.007, which is the
fraction of mass converted to energy during the fusion of Hydrogen.

These last two equations, in cgs units, result in the following relation
between PAH emission at 11.3~{\mm} and starburst luminosities:

\begin{equation}
    L_{\rm SB} ({\rm erg\,\,s}^{-1}) =  \varepsilon c^2 2.2\times 10^{-17}
    L_{\rm PAH}\,\,({\rm erg\,\,s}^{-1}).
\label{sbnuc}
\end{equation}

\noindent Starburst luminosities, derived according to equation \ref{sbnuc},
for all the galaxies in the sample are presented in table \ref{tab.phot}.

It is important to point out that of all the seven galaxies in which we
detected circumnuclear star formation, 4 have already been hinted at having
such a characteristic by previous studies, namely ESO~138-G001, NGC~253,
NGC~2273 and NGC~5506 \citep{CidFernandes2005, Engelbracht1998, Mulchaey1996,
Oliva1999}. Based on the assumption that MIR observations penetrate deeper into
the dusty nuclear environment, one should expect such cases. In the opposite
direction, NGC~6240 and NGC~7469 have previous detections of circumnuclear star
formation in the optical, but we find no evidence for it in our analysis. This
could be either due to the detection limit imposed here or, in the case of
NGC~7469, because the star formation structure seen by previous authors is
outside the pupil used to isolate the nuclear source
\citep[see][]{Soifer2003}.

The SFR estimates presented here for the circumnuclear region of
some of the galaxies are considerably smaller than previous estimates for the
entire galaxy found in the literature. In the case of NGC~6240
\citet{Howell2010} report a SFR of {$148.44\,\,{\rm M}_\odot\,\,{\rm yr}^{-1}$}
using \emph{Spitzer} $L_{\rm IR}$ and GALEX FUV measurements, with the caveat
that some AGN contamination may be present. The most likely explanation for the
smaller value of SFR presented here is the difference in the apertures,
which is almost a factor of 10 smaller when compared to that of \emph{Spitzer}.

On the other hand NGC~253 and NGC~7469 have reported SFRs for the nuclear
region with spatial resolutions comparable to the ones in this paper, with 
values of {$2.8\pm0.3\,\,{\rm M}_\odot\,\,{\rm yr}^{-1}$} and
{$2.6 - 5.1\,\,{\rm M}_\odot\,\,{\rm yr}^{-1}$} respectively
\citep{Ott2005, Davies2007}. Our estimates for these targets are 1.1 dex and
0.6 dex smaller, therefore justifying the conservative uncertainty of 1 dex
mentioned above.



\section{Discussion}


A number of theoretical studies have shown that the interstellar environment
in which SMBH accretion is observed also favour star formation in the
circumnuclear region. For instance, \citet{Kawakatu2008} using a semi-analytic
model show that given a continuous supply of gas from outer parts of the host
galaxy to the inner 100 pc, AGN luminosity will be correlated to SB luminosity
while the accretion rates are high. In another semi-analytic work
\citet{Neistein2014} show that for SB and accretion events ignited by galaxy
mergers, a correlation between $L_{\rm AGN}$ and SFR is verified when
$L_{\rm AGN} > 10^{42}\,{\rm erg\,s^{-1}}$.

In order to investigate the possible physical connection between the AGN and
the circumnuclear starburst, we compared the SFRs probed by the PAH emission to
the accretion rate of the AGNs. The latter quantity was derived from the X-Ray
2-10~keV luminosity, with bolometric corrections obtained from

\begin{multline}
\log \left[ \frac{L_{12}}{L(2-10~{\rm keV})} \right] = \\
 1.54 + 0.24 L_{12} + 0.012L_{12}^2-0.0015L_{12}^3
\end{multline}

\noindent \citep{Marconi2004}, where $L_{12} = \log(L_{\rm AGN}) - 12$ and
$L_{\rm AGN}$ is the bolometric luminosity in units of $L_\odot$. Numerical
methods were used to solve this transcendental equation, resulting in
correction factors roughly between 7 and 55. Resulting bolometric luminosities
for all the AGNs in our sample are also shown in table \ref{tab.phot}. We would
like to emphasise that the value of $L_{AGN}$ obtained from X-Ray radiation is
only representative of the instantaneous accretion rate of the BH. On the other
hand, the emission from aromatic molecules used to assess SFRs lags 150~Myr
behind the main star formation event. 

%
%

\begin{figure}
\includegraphics[width=\columnwidth]{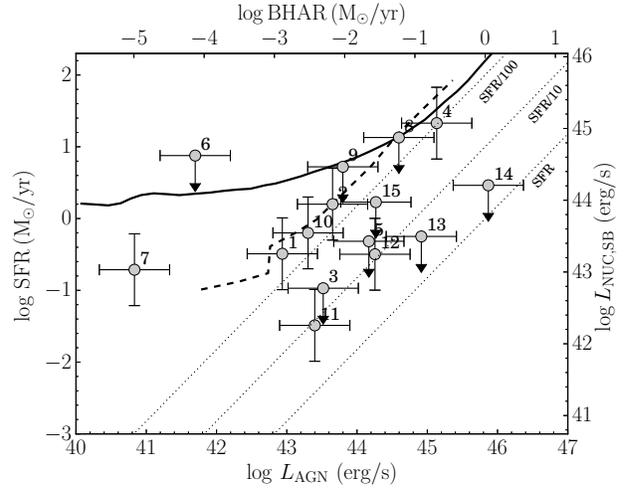}
\caption{Relation between AGN luminosity and the nuclear star formation rate.
    The numbers correspond to the identification on the first column of table
    \ref{tab.phot}. Dotted lines mark accretion rates in terms of the SFR, in
    ratios of BHAR = SFR, BHAR = SFR/10 and BHAR = SFR/100 . The solid line and
    dashed lines are reproductions of the semi-semi-analytic models in
    \citet[][see text]{Neistein2014}.}
\label{fig.agnsfr}
\end{figure}

The points in figure \ref{fig.agnsfr} show the SFRs of circumnuclear
regions
versus the bolometric luminosities of the AGNs. In the same graph we compare
our results with theoretical predictions from \citet{Neistein2014} by
overplotting the average SFRs and $L_{\rm AGN}$ lines from figure 3 in the same
paper. The solid line represents the average SFR for each $L_{\rm AGN}$ bin in
\citet{Neistein2014} models, while the dashed line marks the average $L_{\rm
AGN}$ for each SFR bin. We also note that the higher luminosity AGN in our
sample ($L_{\rm AGN} > 10^{42}\,\,({\rm erg/s^{-1}})$) are mainly located in
the region these authors claim to be occupied by objects in the early stages of
the starburst. Observational evidence from longer wavelengths also show a
similar trend for  high luminosity AGNs. For instance, \citet{Rosario2012}
using 60~{\mm} data from the Herschell space telescope, found that local AGN
luminosities correlate with SFR for
$L_{\rm AGN} > 10^{44}\,\,({\rm erg/s^{-1}})$, thus agreeing with our findings.

Through the use of hydrodynamical simulations \citet{Hopkins2010} have
predicted correlations between black hole accretion rates (BHAR) and SFR. They
conclude that the SFRs are close to BHARs when considering only the nuclear
region (${\rm R} < 10\,{\rm pc}$). The spread in the correlation increases as
larger radii are considered, at the same time that BHARs correspond to smaller
fractions of SFRs. In our analysis we have not compensated the SFRs for the
different projected areas they represent because the majority of our sources is
unresolved. As a result, some of the dispersion in  figure \ref{fig.agnsfr} can
be directly linked to comparing different proportions of the host galaxy.

In figure \ref{fig.agnsfr} we also indicate the corresponding $L_{\rm SB}$ for
the calculated SFRs, meaning the luminosity due exclusively to young stars. The
most striking feature is that there are no galaxies where $L_{\rm AGN} > 100
\times L_{\rm SB}$. In other words, all the AGN harbouring galaxies in our
sample, even the most energetic ones, have starbursts radiating at least 1\% of
the energy from the central engine. At the same time, the high luminosity AGNs
have circumnuclear starbursts that at most match the energy output of the
central source. Moreover we find that two of the seven well constrained
galaxies lie within uncertainty limits of the 1-to-1 line, meaning that a
significant fraction of these sources have nearly as much energy coming from
the AGN as from the circumnuclear starburst. This reinforces the importance of
isolating the AGN emission from star formation at the MIR.

Concerning the low luminosity AGNs, there are is one galaxy which
has more than one hundred times more energy coming from
young stars than from the AGN. This
is in agreement with recent studies showing the huge importance of the
circumnuclear star forming components at all wavelengths to understand LLAGN
\citep[e.g. ][]{GonzalezMartin2014}. Conversely, we find no examples of low
luminosity AGNs with $L_{\rm SB} << L_{\rm AGN}$. Regarding the relationship
between the luminosities from the different phenomena, our data agrees with the
theoretical predictions of \citet{Kawakatu2008} and \citet{Wutschik2013}, which
argue that low luminosity AGNs should not display a correlation between
circumnuclear star formation and AGN activity. According to these authors, this
lack of correlation represents a stage where the circumnuclear disk has become
gravitationally stable, thus ceasing the accretion by the SMBH, but still has
enough gas to form stars.

\citet{DiamondStanic2012} have found a positive correlation between BHAR and
the SFR in radii averaging to 300 pc, throughout their sample, using the
[O {\sc iv}] as an indicator of the former. In that study, only two of the four
points with $L_{\rm AGN} < 42\,\,{\rm erg\,s^{-1}}$ fall within the
correlation, leaving a further two points below it. This result contrasts with
our analysis in the sense that we would expect to see higher circumnuclear star
formation in the low luminosity regime. Using spectra rather than images,
\citet{Esquej2014} arrived at circumnuclear star formation detection rates
similar to ours, almost half of their sample of 23 AGNs. We have four objects
in common with their work, and our estimates for 11.3{\mm} PAH luminosity agree
with theirs within the uncertainties.


\section{Conclusions}

We analysed MIR images of a sample of 15 AGNs in two adjacent filters (PAH2 and
Si5 in the CanariCam, and PAH2 and PAH2\_2 at VISIR) to study the occurrence of
circumnuclear star formation via the 11.3~{\mm} PAH emission band. Three of
these 15 galaxies were observed with GTC/CanariCam, and the remaining
12 with ESO/VISIR, with images taken from the atlas published by
\citet{Asmus2014}. We have also presented new high resolution MIR images for
NGC~2146, which was eventually left out of the remaining analysis due to the
lack of a clearly detectable nucleus. Our main results are as follows. 

\begin{itemize}
\item{
    Circumnuclear star formation, at distances as low as 20 pc from the
    nucleus, was detected in seven out of the 15 galaxies. Among these
    seven galaxies, four show unresolved emission, NGC~253 has clearly
    identifiable star formation nodes, and
    ESO~005-G004 and NGC~2273
    have not been detected in the image subtraction.}
\item{
    The luminosity from the circumnuclear starburst correlates with the
    bolometric luminosity of the central engine only for AGNs with luminosity
    higher than $10^{42}\,\,{\rm erg \,\,s}^{-1}$. In the lower luminosity
    regime, we find that the radiative energy output due to star formation
    tends to be higher than the central engine.}
\end{itemize}

\section*{Acknowledgements}

DRD thanks CNPq and CAPES for partially supporting this research through
fellowships. RR thanks CNPq and CAPES. Partially based on observations made
with the Gran Telescopio Canarias (GTC), installed at the Spanish Observatorio
del Roque de los Muchachos of the Instituto de Astrof\'isica de Canarias, in
the island of La Palma. This research has been partially supported by the
Spanish Ministry of Economy and Competitiveness (MINECO) under the grant
(project ref. AYA 2012-39168-C03-01). OGM acknowledges the Juan de la Cierva
fellowship. 

\label{lastpage}

\bibliographystyle{mn2e}
\bibliography{library}{}

\appendix

\section{Estimating total intensity of the PAH band}
\label{sec.diff2flux}

In this section we will discuss the central issue concerning the determination
of SFR with the available data, which mainly consists of disentangling PAH and
continuum emission with two overlapping filters. Let us assume that the
specific portion of the spectrum in which we are interested, can be well
represented by a continuum plus the PAH emission band

\begin{equation}
    S(\lambda) = C(\lambda) + I(\lambda).
\label{spectrum}
\end{equation}

\begin{figure}
\includegraphics[width=\columnwidth]{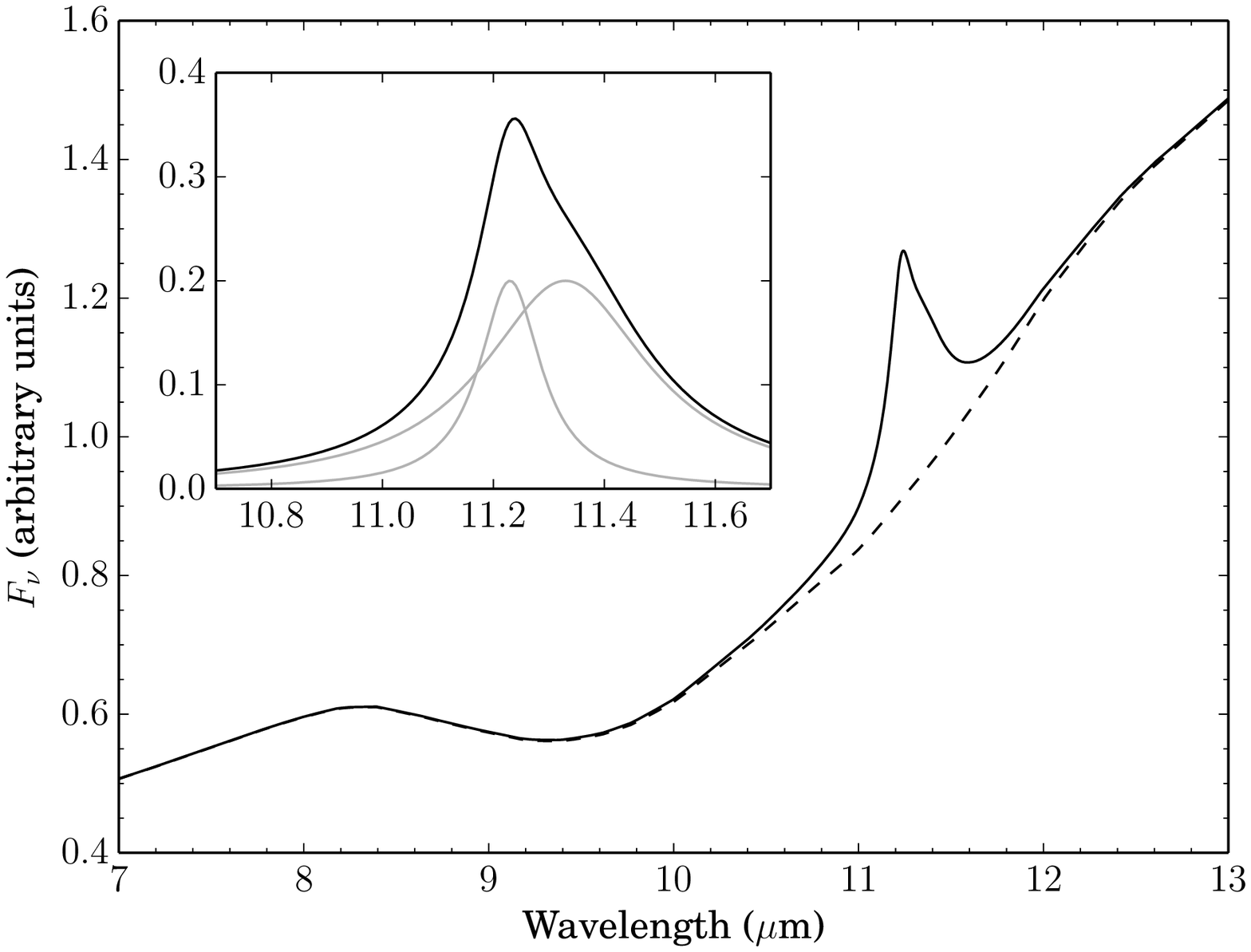}
\caption{
    Example of simulated spectrum used to study the effects of redshift and
    filter subtraction on the estimate of PAH emission. The spectrum that would
    be observed (\emph{solid black line}) is the result of the sum of a
    continuum emission absorbed by silicate grains (\emph{dashed line}) and the
    emission from PAH molecules. The smaller plot window highlights the
    composition of the 11.3~{\mm} by two distinct Drude profiles (\emph{grey
    lines}.)}
\label{fig.simspec}
\end{figure}

\noindent The later is described here by the sum of two Drude profiles
parametrised by their central intensities as in \citet{Smith2007a}:

\begin{equation}
    I_\nu = \frac{b_r \gamma_r}{\left( \lambda/\lambda_r -
        \lambda_r/\lambda \right)^2 + \gamma_r^2}.
\label{drude}
\end{equation}

\noindent where $b_r$ is the central intensity, $\gamma_r$ is the fractional
full width at half maximum ${\rm FWHM}/\lambda_r$ and $\lambda_r$ is the
central wavelength \citep{Smith2007a}. The values of $\lambda_r$ and $\gamma_r$
that define the first component are 11.22~{\mm} and 0.012, respectively, while
11.33~{\mm} and 0.032 represent the second component. Just as in section
\ref{sec.pahmaps}, the spectral decomposition code {\sc pahfit}
\citep{Smith2007a} was employed to obtain individual models for the continuum
of each galaxy based on \emph{Spitzer}/IRS spectra. Figure \ref{fig.simspec}
shows an example of a modelled spectrum, with the average continuum of our
sample. In this example, and in all subsequent analysis, the central intensity
of both Drude profiles was kept equal. Tests were performed to evaluate the
impact of different ratios of central intensities, and no large differences
were found.

\begin{figure}
\includegraphics[width=\columnwidth]{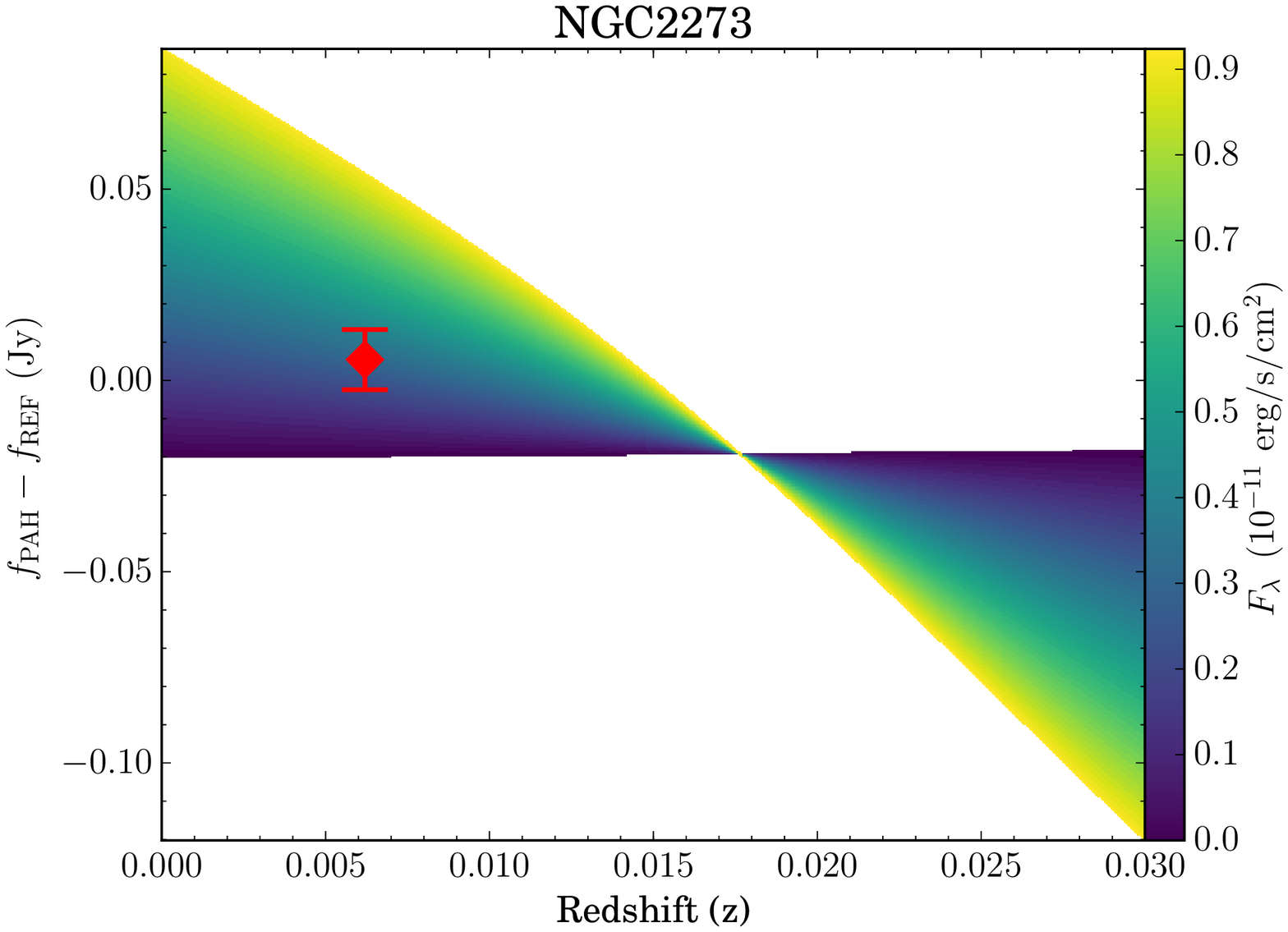}
\caption{
    Analysis of the effects of redshift and total intensity on the 
    difference between the filters on CanariCam. At a redshift of 0.017
    the reference filter begins to have a higher flux than the emission filter
    even for the highest intensities allowed in the models.}
\label{fig.d2f2273}
\end{figure}

\begin{figure}
\includegraphics[width=\columnwidth]{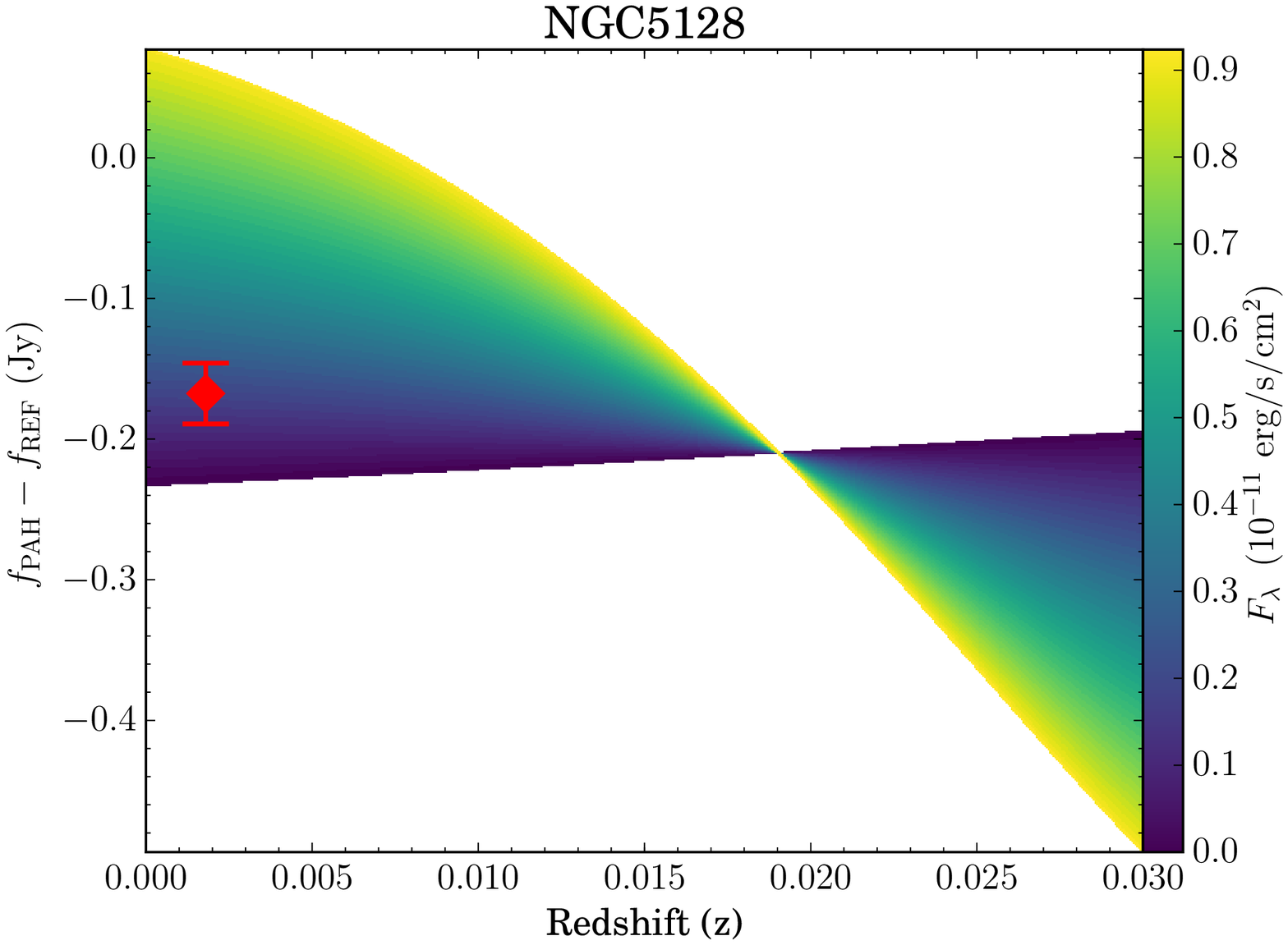}
\caption{
    Same as figure \ref{fig.d2f2273} for the galaxy NGC~5128}
\label{fig.d2f5128}
\end{figure}

\begin{figure}
\includegraphics[width=\columnwidth]{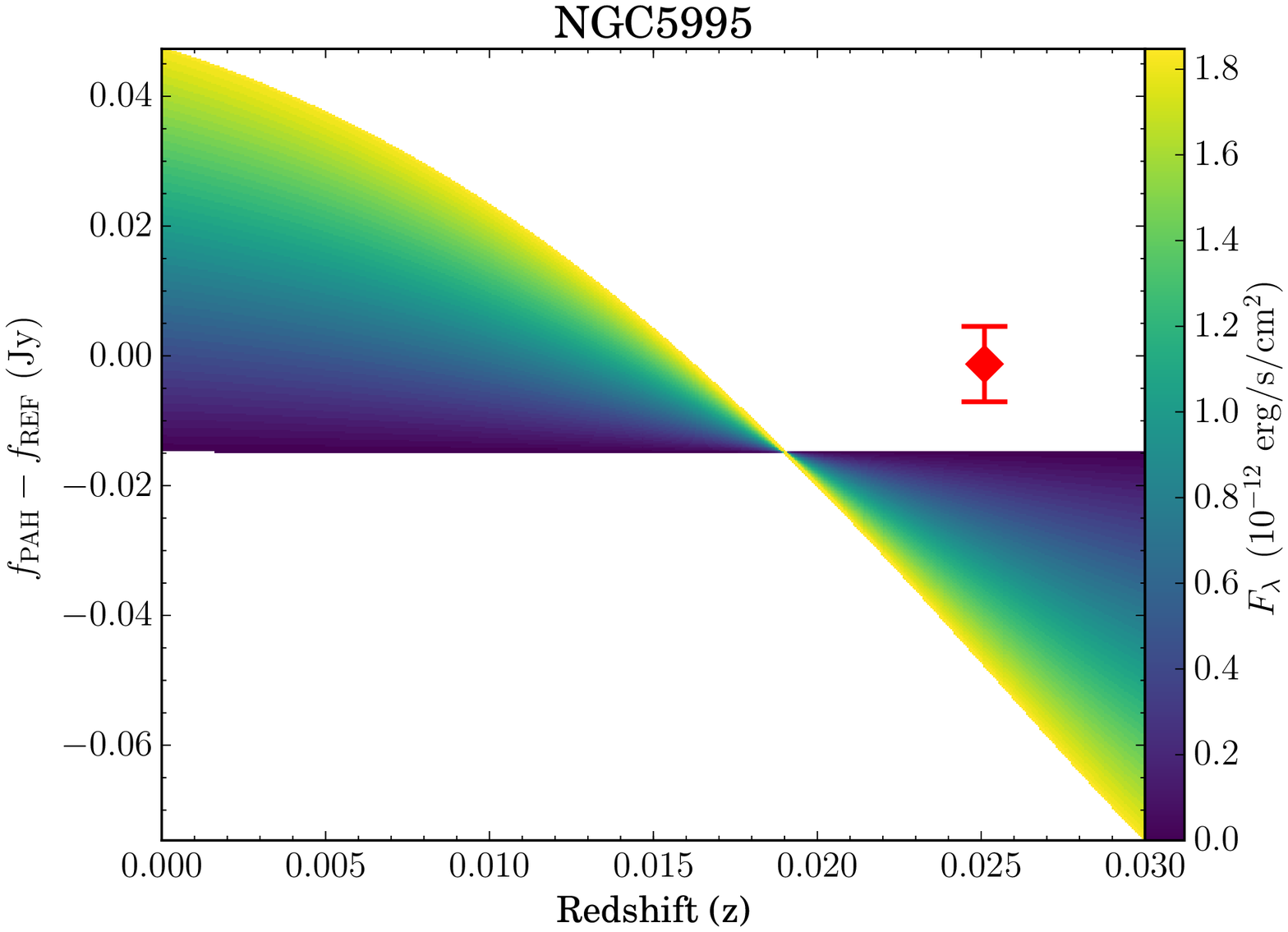}
\caption{
    Same as figure \ref{fig.d2f2273} for the galaxy NGC~5995}
\label{fig.d2f5995}
\end{figure}

The total flux $f_n$ measured in each filter is essentially a weighted average
of the spectrum, with the weight for each wavelength given by the filter's
transmission function. Thus

\begin{equation}
    f_n = \int F_n(\lambda)S(\lambda)\,{\rm d}\lambda.
\label{filterflux}
\end{equation}

\noindent where $F_n$ is the normalised transmission curve. It is easy to see
that, within the framework of this model, the difference in flux between
filters, that is $f_{\rm PAH} - f_{\rm REF}$, can be written as
function of only the central intensity of the PAH emission profiles, and the
redshift of the galaxy: 

\begin{multline}
D(b_r,z) = \int_{\lambda_0}^{\lambda_1} \left[ C_\nu(\lambda,z)
    + I_\nu(b_r,\lambda,z) \right] \\
    \times \left[ F_1(\lambda) - F_2 (\lambda) \right]\,{\rm d}\lambda
\label{diff}
\end{multline}

\noindent where $C_\nu(\lambda,z)$ is the continuum emission,
$I_\nu(b_r,\lambda,z)$ is the PAH emission in the 11.3~{\mm} band, and
$F_1(\lambda)$ and $F_2(\lambda)$ are the normalised transmission functions for
the PAH and reference filters respectively. The total intensity of the PAH band
is given by the sum of the total intensity of each component, which in turn is
the integral of equation \ref{drude}: 

\begin{equation}
    F_{\rm PAH} = \frac{\pi c}{2} \left( \frac{b_1 \gamma_1}{\lambda_1} +
        \frac{b_2 \gamma_2}{\lambda_2} \right)
\label{totalintens}
\end{equation}

\noindent where the sub indexes represent the different components. Therefore,
once equation \ref{diff} is solved for a particular central intensity $b_r$,
and assuming the redshift $z$ is known, we can use the result in equation
\ref{totalintens} to calculate the total flux in the 11.3~{\mm} PAH band.
Numerical methods were employed to investigate the solutions of \ref{diff},
and figures \ref{fig.d2f2273} through \ref{fig.d2f5995} show three distinct
examples of typical outcomes.

In figure \ref{fig.d2f2273} we show a map of solutions in the plane
$f_{\rm PAH} - f_{\rm REF}$, which we assume to be equal to
$D(b_r,z)\,{\rm vs.}\,z$, for the inferred continuum of NGC~2273 and the
CanariCam filters. The red diamond in this figure represents the measured
filter difference, along with 1-$\sigma$ error bars. The colour scale
represents the total flux of the simulated PAH band, from deep blue for
no emission to bright green for the maximum simulated emission. The latter
is completely arbitrary and was chosen for readability purposes.
In the case of NGC~2273, the almost flat slope of the continuum has no clear
impact on the difference between fluxes, for any of the sampled redshift
coordinates. Nevertheless, it is already clear that even a modestly negative
value of $D$ is still compatible with a PAH emission of about
$0.2\,{\rm erg/s/cm^2}$. At a redshift of 0.017 the situation is reversed,
because the reference filter ($f_{\rm REF}$) begins to sample more of the PAH
band than the original PAH filter. Thus, for the same continuum shape and for
the same filters, the PAH emission of a galaxy with redshift $z > 0.017$,
would cause $f_{\rm PAH} - f_{\rm REF}$ to be lower than -0.2 Jy.

Figure \ref{fig.d2f5128} shows the example of NGC~5128, which has a 
notably steeper continuum (see fig. \ref{fig.spectra2}). At the redshift of
this galaxy the flux of PAH band would have to be almost
$8 \times 10^{-10}\,\,{\rm erg/s/cm^2}$ for the difference between fluxes to
be zero. The negative flux seen in the image subtraction of section
\ref{sec.pahmaps} is therefore completely compatible with PAH emission,
although the former alone is not enough to warrant an emission detection.
The slowly rise in the minimum value of $f_{\rm PAH} - f_{\rm REF}$ that can
be seen in figure \ref{fig.d2f5128} is caused by the change in continuum slope
as one moves towards longer wavelengths, leaving the silicate absorption band
at 9.7 $\umu$m.

Finally we discuss the case of the NGC~5995, the galaxy with the highest
redshift in our sample. This galaxy is well beyond the threshold where
the reference filter has more of the PAH band than the original PAH filter,
which in the case of VISIR happens at $z = 0.019$. A positive detection would
therefore be characterised by $f_{\rm PAH} - f_{\rm REF} \le -0.015$ Jy.
However the photometric measurements show practically no difference between
the fluxes in both filters, thus placing NGC~5995 in a region which would 
imply a negative PAH flux, or a more likely misrepresentation of the
underlying continuum. We classify such cases as non-detections and
report PAH fluxes and derived quantities as upper limits, based on the
photometric uncertainty. As for the reasons that might have thus affected the
continuum, the most probable is a simple overestimation of its declivity.
Another possibility would be the presence of silicate emission at the 9.7
$\umu$m band, rather than the more usual absorption. These emission features
are a common occurrence among QSOs \citep[e.g. ][]{Hao2005}, and have also been
identified in many local AGN
\citep{Sturm2005, Thompson2009, Mason2012, RuschelDutra2014}, and are
predicted by radiative transfer models based on clumpy tori even for
Seyfert 2's \citep{Nenkova2008}.


\end{document}